\documentclass[10pt,letterpaper]{article}
\usepackage{opex3}
\usepackage{graphicx}
\usepackage{amssymb,amsmath,amsfonts}

\begin{document}

\title {Mid-infrared guided optics: a perspective for astronomical instruments}

\author{Lucas Labadie,$^{\,1\ast}$ and Oswald Wallner$^{\, 2}$}
\address{$^{1 \,}$Max-Planck Institut fuer Astronomie, Koenigstuhl 17, D-69117 Heidelberg, Germany \\ $^{2 \,}$EADS Astrium GmbH, 88039 Friedrichshafen, Germany \\
$^{\ast\,}$Corresponding author:\,{\footnotesize {\it \textcolor{blue}{\underline{labadie@mpia.de}} \rm } \normalsize}}




\begin{abstract}
Research activities during the last decade have shown the strong potential of photonic devices to greatly simplify ground based and space borne astronomical instruments and to improve their performance. We focus specifically on the mid-infrared wavelength regime (about 5--20\,$\mu$m), a spectral range offering access to warm objects (about 300 K) and to spectral features that can be interpreted as signatures for biological activity (e.g. water, ozone, carbon dioxide). We review the relevant research activities aiming at the development of single-mode guided optics and the corresponding manufacturing technologies. We evaluate the experimentally achieved performance and compare it with the performance requirements for applications in various fields of astronomy. Our goal is to show a perspective for future astronomical instruments based on mid-infrared photonic devices.
\end{abstract}



\section{Introduction}

\subsection{Scientific drivers for mid-infrared astronomy}

Any celestial object with a non-zero temperature emits infrared radiation (i.e. heat). Because the wavelength at which the object most intensively radiates depends on the temperature, observing in the spectral range that goes from 1\,$\mu$m to 1000\,$\mu$m gives access to objects at a wide range of temperatures from a few tens of degrees Kelvin to several thousands.
Although spectral boundaries are not firmly set, the region from 1\,$\mu$m to 5\,$\mu$m is defined as the near-infrared, from 5\,$\mu$m to $\sim$\,30\,$\mu$m we refer to the mid-infrared domain  (or {\it thermal} infrared), while the far-infrared extends up to 1000\,$\mu$m.
The mid-infrared becomes a highly interesting observing regime, where warm objects with temperatures of $\sim$100--600\,K can be probed. This corresponds to a temparature range where physical conditions for liquid water are encountered. It is also the range of peak emission of telluric planets around solar-type stars and therefore an appropriate domain for detecting the self-emission of nearby exoplanets. Furthermore, bio-markers such as water, CO$_{\rm 2}$ or O$_{\rm 3}$ (ozone) can be revealed through low-resolution spectroscopic observations.
Mid-infrared sources have become unquestionably a powerful source of information for relatively new fields like astrobiology and astrochemistry.

\subsection{Photonics, a recent technology for astronomy}

New technologies have constantly played a key role in improving the performance of astronomical instruments.
With technology advancing at a rapid rate, future projects like the {\it E-ELT} (European Extremely Large Telescope), multi-aperture imaging or space-based infrared interferometry require us to fully reconsider the approach adopted so far for classical instruments. This can be justified by the high level of complexity of the classical solutions, the requirement of unprecedent performance or the need for reducing physically the size of future instruments. In this context, photonics can undoubtedly help to face some of this challenge. In the last decade, there has been an increasing and innovative use of photonic devices in astronomy.
Optical fibers, the simplest photonics elements, are now commonly used for fiber-fed spectrographs in the visible. In the specific area of stellar interferometry, single-mode optical fibers are valuable for their spatial filtering capabilities, which help improving the accuracy on the interferometric visibilities \cite{Foresto}. Fibers are currently used on various interferometers like CHARA \cite{Alister} or AMBER/VLTI \cite{Petrov} and will be implemented in future instruments to interconnect telescopes separated by a few hundred meters distance \cite{Perrin}. Beside optical fibers, Integrated optics (IO) is very promising for next generation multi-aperture interferometers 
since it offers the possibility of combining several beams in a single chip, leading to obvious advantages in terms of space and stability. This concept was implemented in the past on the IOTA interferometer \cite{Berger, Schloerb}, and is foreseen to be part of the VLTI second generation instrumentation \cite{Malbet, Eisenhauer}. This is the prelude to a valuable miniaturization of the interferometric instrumentation, possibly extending it in the future to other instrument concepts like integrated spectrograph \cite{Bland-Hawthorn, LeCoarer}.
Photonic devices could also be helpful in the context of space-based experiments, where simple designs, and reliable and stable components are absolutely necessary. This points are important in particular for future interferometric missions like {\it Darwin/TPF}, {\it Pegase} or {\it SIM}.


\indent Photonics has several strong advantages from the instrumental point-of-view, which justifies the extension of the concept towards new wavelengths. Some of these aspects which have been described in detail in \cite{Malbet1999} deserve our interest.
The compactness offered by photonics-based components is a major advantage in the context of large telescope instrumentation and multi-aperture interferometry: {\it integrated} features allows us to save space and weight, and at the same time reduces the overall instrumental complexity. Photonics components can be placed in direct contact with a cryostat window or a detector \cite{LeCoarer}; They allow flux transportation in a much easier way than long optical trains; They permit faster and simpler instrument upgrade when dealing with optical chips; {\it single-mode} fibers and integrated optics allow very accurate beam cleaning, thanks to spatial and modal filtering, which is an essential aspect of aperture synthesis and nulling interferometers \cite{Ollivier1997, Mennesson2002}; When the waveguide is placed in a chilled enclosure, its limited numerical aperture acts as a field stop, avoiding straylight entering the system from outside the acceptance cone. This is an advantage when operating in the infrared, where thermal background from nearby warm optics is very high.

\indent From these different points, photonics appears very attractive in the context of instrument miniaturization. However, the strong telecom background of photonics has concentrated the technological effort in few narrow bands of the near-infrared spectrum, where fully mature technology is available. When considering longer wavelengths of the {\it astronomical} mid-infrared spectrum, photonics suffers from a significant technological gap, which results from a weak mastering of the materials production and manufacturing processes at these wavelengths. Promising efforts, mainly driven by the potential applications of photonics in fields like astronomy, chemistry and biophysics, are being made to reduce the gap. The increasing use of photonic devices for astronomical instrumentation has led to the neologism {\it astrophotonics}.



\section{Materials and concepts for mid-infrared photonics}

\subsection{Accessing the 10-$\mu$m spectral range: materials selection}

The selection criteria for materials are based on the transparency window and on the corresponding intrinsic and extrinsic losses within this window. The window borders are determined by absorption, both on the visible and mid-infrared part of the spectrum. The intrinsic losses depend only on the material structure and properties and represent the minimum expected losses. However, the presence of impurities and defects further degrades the transmission and these losses are generally more important. We then talk about extrinsic losses. In the particular case of silica in the telecommunications industry, the refinement of the manufacturing technology over several years has led to high level of purity and quality, which makes the material losses close to the intrinsic levels. The physical explanation for intrinsic losses are found in material electronic transitions, phonon interactions and free-carrier effects. 
The interested reader can refer to well-established reference books on optical and infrared materials \cite{Klocek}.
The infrared materials that have been tested so far in the context of dielectric photonics devices are chalcogenide glasses, silver-halide glasses and zinc selenide components. Most of these are also transparent in the visible. This has to be considered in case the resulting components have to operate in an environment non-shielded from optical radiation. On the other hand, this offers an advantage for alignment. All these materials are transparent at least up to 11-12\,$\mu$m. Depending on the chemical composition and the stoichiometry, the transmission range can be extended up to 18--20\,$\mu$m for the chalcogenide and zinc selenide glasses, or even beyond for silver halide glasses.
In addition to their intrinsic properties, other factors may influence the optical behavior of these materials. These are:\vskip 0.25cm
\noindent -- mechanical stress: most infrared materials show a certain fragility and high sensitivity to shocks, which may complicate the manufacturing and polishing processes.\vskip 0.1cm
\noindent -- chemical stability: some materials can absorb water vapor, which further affects their transparency. In some cases, vacuum operation is required.\vskip 0.1cm
\noindent -- thermal stress: while fibers are sensitive to temperature, integrated optics are generally more stable with regard to thermal constraints \cite{Berger2000}.
\vskip 0.1cm
\noindent -- spatial homogeneity: when a large surface is required to manufacture complex devices, it is important to ensure a good degree of homogeneity of the optical properties over the whole sample. Underestimating this aspect might bring in differential effects (dispersion etc...) that must then be carefully controlled.\vskip 0.25cm

Infrared glasses are typically classified in amorphous and crystalline forms, depending on the regular (crystalline) or irregular (amorphous) arrangement of the atomic structure. \\ 
\indent Amorphous materials include oxide glasses (e.g. SiO$_{\rm 2}$) which present strong absorption in the mid-infrared due to OH vibration modes. Other members of this group are
chalcogenide glasses (based on S, Se and Te). The transmission of these materials can be extended beyond 12\,$\mu$m by the addition of heavy compounds (Ag, Te).\\
\indent Crystalline materials show a regular pattern in their structure (e.g. Quartz, the crystalline form of SiO$_{\rm 2}$). Silver halides -- bromide or chloride -- are also crystalline materials well known in mid-infrared applications. Semiconductors, also classified in this category, are attractive materials due to their extended transmission range in the infrared. \\
So far, 
chalcogenide and silver halide glasses are the two main materials that have been investigated to cover the mid-infrared range. Typical chalcogenide glasses are As$_{\rm 2}$Se$_{\rm 3}$3 and As$_{\rm 2}$S$_{\rm 3}$, which present good transparency up to $\sim$12--16\,$\mu$m. Compositions including Germanium (Ge) or Tellurium (Te) can extend the transmission range up to 20\,$\mu$m. Silver halide glasses have excellent transparency up to and beyond 20\,$\mu$m.
In terms of materials properties, silver halide glasses may encounter corrosive risks if in contact with metals. Furthermore, they present photosensitive properties, which require particular care if this effect is undesirable. On the other hand, the material photosensitivity can also been exploited for controlled laser writing of waveguides (see Sect.~\ref{IO}). 

\subsection{Waveguide concepts}



\subsubsection{Optical fibers and integrated optics}

These types of dielectric waveguide represent a natural choice for
developing guided optics for the mid-infrared, since they
represent a spectral evolution of near-infrared concepts. We raise
here only those issues which are important for the rest of the
paper. For a detailed study of optical fibers and integrated
optics principles, the reader can refer to classical references
\cite{snyder1983, Lee}.
Several refractive index profile can be
considered when designing optical fibers: single-step index
profile, double-step index profile, and smooth profile. The
opto-geometrical parameters of the fiber (core and cladding size,
refractive index difference $\Delta$n) have a strong influence on
the transmission range, the field confinement (strong or weak) or
the numerical aperture. Most relevant for some astronomical
applications are single-mode fibers. An important parameter to
measure modal behavior is the normalized frequency of the fiber
$V=2\pi\rho/\lambda(n_{co}^2-n_{cl}^2)^{1/2}$ with core radius
$\rho$ and wavelength $\lambda$ to be compared to the first order
cut-off frequency $V_c=2.405$. Depending on the index profile of
the fiber, the single-mode bandwidth can be theoretically infinite
(single-step index and smooth profiles) or finite (double-step
index fibers under certain conditions on the cladding index).
As for classical fibers, {\it planar} dielectric integrated optics
confines the light by total internal reflection. The main
difference with fibers is on the waveguide geometry and on a stiff
and fixed arrangement of the waveguides on the substrate surface.
The basic brick is the slab waveguide, whose modal properties are
governed by the dispersion equation of the slab waveguide
\cite{Labadie2006}.
An integrated optics chip can gather  within a small glass surface a high number of optical functions. It becomes a very powerful solution for interconnecting interferometrically several telescopes and obtaining accurate phase-shifted outputs for measuring visibilities \cite{Benisty2006}.
Non-linear crystals such as Lithium Niobate (LiNbO$_{\rm 3}$) are used to build active components, such as electro-optic modulators with high accuracy phase control. The combination of integrated optics beam combiners and dispersing elements (e.g. AWG or Arrayed Waveguide Gratings) could lead to promising end-to-end integrated instruments with unprecedent mechanical stability.
In practice, as for many photonic device, integrated optics chips present a certain number of limitations on their wide spread use in astronomy.
The effective bandwidth of operation is generally narrow and the optical functions are highly wavelength dependent. Cross-talk effects between different channels can degrade the useful signal, although this can reduced by a careful design of the chip.


\subsubsection{Hollow waveguides and photonic-crystal fibers}\label{HW1}

\begin{figure}[t]
\centering\small
\begin{tabular}{r c c c}
\multicolumn{2}{c}{{\bf Hollow waveguides}} &
\multicolumn{2}{c}{{\bf Photonic crystal fibers}}  \\ [0.25cm]
& & index-guiding & PBG-guiding \\[0.2cm]
\centering\includegraphics[width=0.26\textwidth]{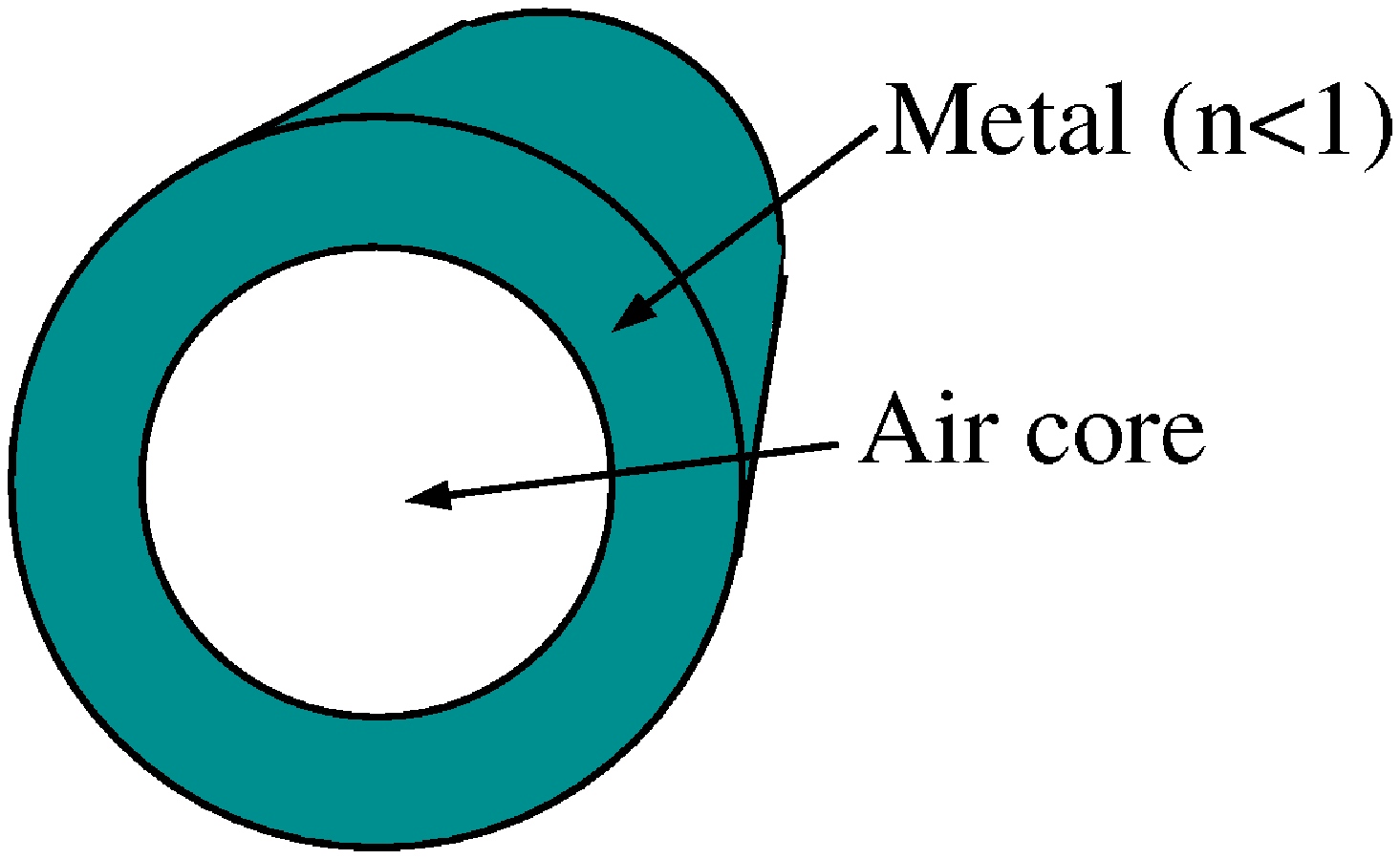} &
\centering\includegraphics[width=0.29\textwidth]{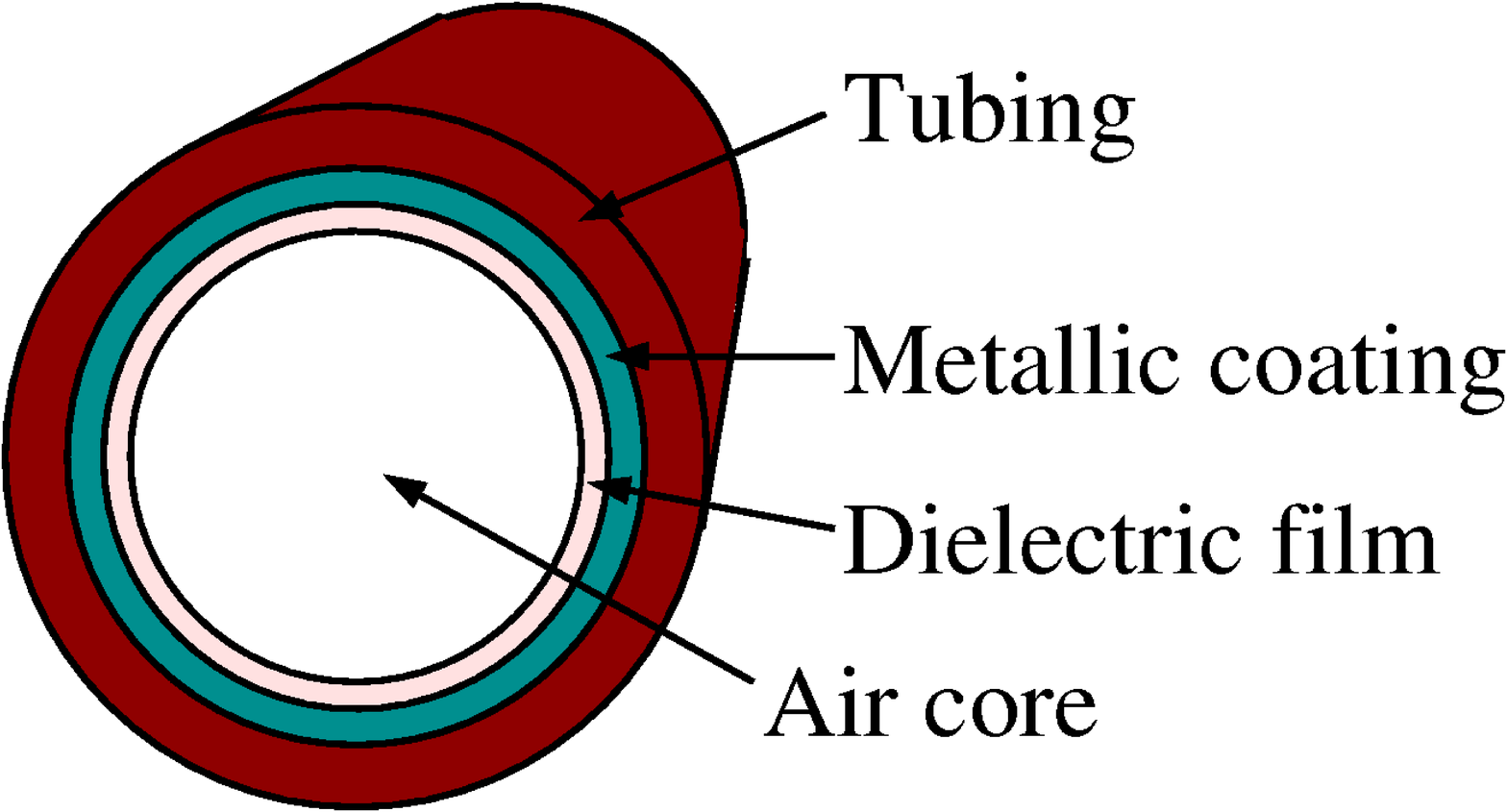} &
\centering\includegraphics[width=0.17\textwidth]{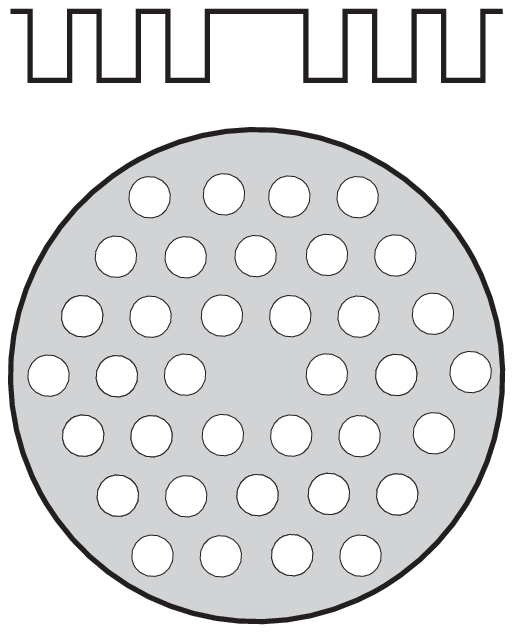} &
\centering\includegraphics[width=0.17\textwidth]{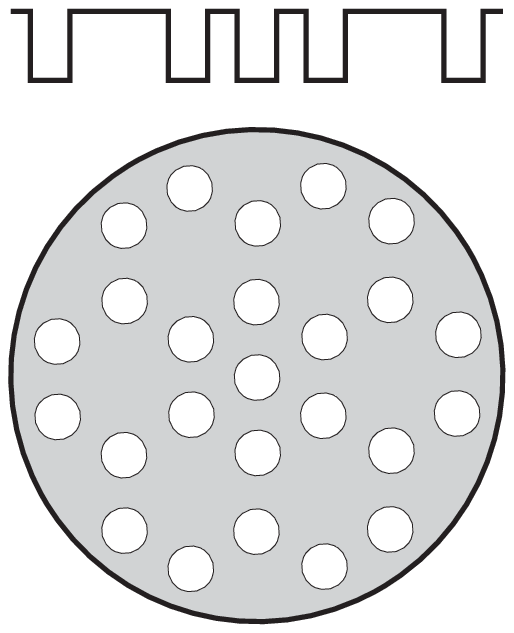}
\end{tabular}
\caption{Left:\, principles of hollow metallic waveguides and hollow glass waveguides. In the HGW design, the tubing can be glass or a plastic polymer. Right:\, refractive index profiles of index-guiding and photonic bandgap (PBG) guiding PCF.
\label{fig1}}
\end{figure}

Hollow waveguides were initially proposed as an alternative to
dielectric fibers because of the lack of core infrared materials
with suitable structural properties. In hollow waveguides,
infrared radiation propagates in ideal low-loss media like air or
vacuum thanks to specular reflections on the metal-coated walls of
the structure. Following the first experimental demonstration  of
this concept \cite{Garmire1976, Garmire1980}, a large variety of
hollow waveguides has been developed based on the air-core
multi-layer structure. Two main class of hollow waveguides can be
distinguished (see Fig.~\ref{fig1}): hollow metallic waveguides
(HMW) are made of metallic tubes or metallic layers deposited
inside a micromachined structure. Because of the specular nature
of the electromagnetic wave reflection, there is no limiting
entrance angle as in dielectric fibres. Imperfect reflectivity and
roughness of the coating are responsible for propagation losses. A
second class is hollow glass waveguides (HGW) or hollow plastic
waveguides (HPW) for which glass -- or plastic -- tubing is used
with an internal multi-layer dielectric-metallic coating. In this
case, leaky modes are confined by the underlying metallic coating.
HGW present lower losses over a broader spectral range compared to
HMW, due to a smoother inner wall. Circular waveguides with core
diameter from 200\,$\mu$m to 1000\,$\mu$m are commonly found, but
they are highly multimode and not compliant with the single-mode
property required in some astronomical applications. Because of
the importance of single-mode behavior for several applications,
some authors have theoretically investigated the modal behavior in
hollow waveguides with various geometries and emphasized the
conditions for single-mode behavior \cite{Vermeulen1991,
Bernini2002}. This lead to the identification of quasi single-mode
circular waveguides and strictly single-mode rectangular
waveguides. Such single-mode waveguides are quite lossy with
attenuation around $\sim$30\,dB/m, which limits their use to short
propagation lengths. The typical cross-section dimensions of {\it
single-mode} waveguides is of the order of one to two wavelengths,
i.e. 10 to 20\,$\mu$m in the mid-infrared range. It is only
recently that micro-machining capabilities permitted the
production of hollow waveguides at this scale, allowing fruitful
synergizes between microwave engineering and infrared photonics.
Photonic crystal fibers (PCF) are made of dielectric structures
with periodically varying refractive index at the wavelength scale
\cite{bjarklev2004}. These structures exhibit photonic bandgaps,
which are frequency intervals within which light propagation is
forbidden in certain directions.
PCF are formed from a dielectric rod containing tubes of a material of lower refractive index or air, arranged in a special pattern and running along the fiber axis.
Two types of photonic crystal fibers may be distinguished: those
with an index-guiding structure and those with a photonic
bandgap-guiding structure (see Fig.~\ref{fig1}). The pattern of
air or low refractive index tubes determines the waveguiding
properties. Single-mode operation is possible for both types of
PCFs , although with different single-mode bandwidth and different
propagation properties:
\noindent -- Index-guiding PCFs
are formed by a not necessarily periodic structure of air or low refractive index tubes at wavelength scale, embedded in some dielectric material. A defect which confines the light and therefore acts as the core is introduced by omitting one tube. Waveguidance can be explained by considering that the cladding region provides (due to the tubes) an effective
index which is smaller than the core index.
Depending on the wavelength, the effective index approaches the material index of the core, if the field is mainly guided within the material, or the index of the tubes, if the field penetrates the cladding. With the effective cladding index approach, the index-guiding PCF can be modeled as a single-step index fiber with an effective normalized frequency of $V_{e\!f\!\!f}=\pi\Lambda/\lambda(n_m^2-n_{e\!f\!\!f}^2)^{1/2}$, where the core diameter is given by the center-to-center spacing $\Lambda$ between two adjacent tubes. In contrast to conventional single-step index fibers, $V_{e\!f\!\!f}$ approaches a constant value for large $\Lambda/\lambda$. This means that by proper choice of the hole diameter to hole spacing ratio $d/\Lambda$, such a fiber can be single-mode for any wavelength \cite{birks1997}.\vskip 0.1cm
\noindent -- Photonic-bandgap (PBG) guiding PCFs are obtained by locally breaking the periodicity of a photonic crystal by introducing a well-defined defect, e.g.\ in the form of an extra tube. The light then is confined and thus guided, provided that the surrounding photonic crystal cladding exhibits a photonic bandgap at the operation wavelength \cite{barkou1999}. A remarkable difference between PBG-guiding PCFs and conventional fibers or index-guiding PCFs is that they allow waveguidance with propagation constants (i.e.\ photonic bandgaps) below the effective cladding index. The photonic bandgap regions, i.e.\ the transmission windows, are determined by the cladding structure only. They are usually very narrow. The behavior of the guided modes within the photonic bandgaps is solely determined by the characteristic of the defect. By varying the size and shape of the defect, the frequency of the guided mode can be positioned precisely within a photonic bandgap. At near-infrared wavelengths, substantial work has led to the implementation of PCF in astronomical applications \cite{vergnole2005, vergnole2006}. \vskip 0.25cm

\section{Overview of development activities}

\subsection{Manufacturing techniques}\label{manufacture}

\subsubsection{Optical fibers}

\paragraph{Chalcogenide glass fibers:}

The manufacturing techniques for optical fibers are determined by the the different physical properties of fiber materials. In the mid-infrared we distinguish between glassy fibers and crystalline fibers. For both types, a preform is created by mechanical combination of the core and cladding material. Glassy fibers are drawn from the preform, whereas crystalline fibers are extruded from the preform.\\
\indent Chalcogenide glass was the first material used to produce
mid-infrared fibers. A two or more-component glass is formed by
combining one or more chalcogene elements such as sulphur,
selenium or tellurium with one or more elements such as arsenic,
germanium, phosphorus, antimony, gallium, aluminum or silicon. The
properties of the glass change drastically with composition. In
general, chalcogenide glasses are very stable, durable, and
insensitive to moisture. A distinctive difference between these
and other mid-infrared fiber glasses is that they do not transmit
well in the whole visible region and that their refractive indices
are quite high. Arsenic-Trisulfide (As$_2$S$_3$) fibers have a
transmission range from $0.7$ to about $12\ \mu m$. Longer
wavelengths can be transmitted by adding heavier elements like
Tellurium, Germanium or Selenium. Chalcogenide glass formation can
be achieved when the chalcogenide elements are melted and quenched
in an evacuated silica glass ampoule with one or more elements as
mentioned above.  After heating in a rocking furnace for longer
than 10 hours at a temperature between $700$ and $900^\circ C$,
the ampoule is quenched in air and annealed to room temperature.
The preform is fabricated by using rod-in-tube (core rod is placed
within the cladding tube) or casting methods (melted core material
is cast into the cladding rod). The two methods in use for fiber
drawing are preform drawing and crucible drawing. In the case of
fibers with no glass cladding or only a polymer cladding,
e.g.\,teflon, the conventional preform drawing method with a tube
furnace and a narrow heat zone within an inert gas atmosphere is
used. Fibers with chalcogenide cladding are drawn in a similar
apparatus. The core rod and cladding tubes are placed into a
silicon muffle furnace, zonally heated, and drawn into a fiber.
Core-cladding fibers can also be produced  by a double-crucible
technique or by modified crucible drawing, a technique offering
the advantages of both crucible drawing and preform drawing
\cite{sanghera1998}. Single-mode fibers made of chalcogenide glass
have been and are currently developed for ESA and NASA in the
framework of the planned missions DARWIN and TPF-I. The fibers are
intended as spatial wavefront filters for the wavelength range
from about $4$ to $12$\,$\mu$m. {\bf Several} independent research
activities demonstrated the feasibility of single-mode fibers:
\noindent-- A team at LESIA (Observatoire de Paris, France) and Le
Verre Fluor\'e (France) developed a single-mode fiber of
chalcogenide glass for spatial wavefront filtering
\cite{borde2003}. The fiber had a core/cladding composition of
As$_2$Se$_3$/Ge$_2$SeTe$_{1.4}$ with a numerical aperture of
$0.15$ and core and cladding diameters of $40$\,$\mu$m and
210\,$\mu$m, respectively.
\noindent -- A team at TNO (The Netherlands) and the University of Rennes (France) developed for ESA several samples of chalcogenide fibers by different preform manufacturing methods, including internal rotational casting, fiber or rod in tube vacuum. Single-mode operation has been demonstrated for a fiber composed of Te-As-Se glass with a core diameter of 22\,$\mu$m, a cladding diameter of 500\,$\mu$m, core and cladding refractive indices of 2.927 and 2.924, and a nominal single-mode cutoff wavelength of $3.7$\,$\mu$m \cite{cheng2005}.
In a second activity the team is analyzing tellurium-based chalcogenide fibers for the use in the wavelength range from about $10$ to $20$\,$\mu$m. Mono-index fiber samples with compositions of Te-Ge-Ga-I, Te-Ge-I and Te-Ge-Se have been produced with diameters in the range of $400$ to $550$\,$\mu$m.
Because the properties of Tellurium-based glass depend on
temperature, the fiber samples show improved transmission at
cryogenic temperatures.
\noindent -- The Naval Research Laboratory
(United States) developed for NASA chalcogenide glass single-mode
fibers for use at wavelengths up to about $11$\,$\mu$m. The fibers
had a core diameter of $23$\,$\mu$m, a cladding diameter of
$127$\,$\mu$m, and core and cladding refractive indices of $2.725$
and $2.714$ \cite{ksendzov2007}.

\paragraph{Silver-Halide Fibers}

Crystalline materials have the advantage of a better long wavelength transmission compared to mid-infrared glasses, but suffer from a difficult fabrication process. The first crystalline fibers were made of hot extruded KRS-5 (Tellurium-Bromide-Iodide). Today, poly-crystalline silver-halide (Silver-Chloride-Bromide) is most widely used. Silver-halide fibers show good transmission up to almost $20\ \mu m$.
Poly-crystalline fibers are made of crystal-like solid solutions of Thallium halides, e.g.\ KRS-5, or silver halides, e.g.\ AgClBr. These materials
offer such properties as ductility, low melting point, and isotropy. Crystalline fibers can be fabricated via plastic deformation by extrusion from a preform. The rod-in-tube preform is realized by a rod of AgClBr as core and a Cl-rich AgClBr crystal as cladding. The preform can also be made by the casting method or by preform growth methods \cite{artiouchenko2000}. The preform is placed in a heated chamber and the fiber is extruded to its final form through a polished die. Typical extrusion temperatures are in the range of $50\%$ to $80\%$ of the melting temperature, which is $457^\circ C$ for AgCl, for example. The pressure within the container ranges form a few to some ten tons per square centimeter.
Similar to the chalcogenide glass fibers, poly-crystalline silver-halide fibers have been and are currently being developed for ESA and NASA.
Two independent research activities demonstrated the feasibility
of single-mode fibers:
\noindent -- A team at Astrium GmbH
(Germany) and ART Photonics (Germany) developed for ESA several
samples of silver-halide fibers with different core/cladding
geometries and different material compositions
\cite{flatscher2006}. To achieve optimum results, extremely
homogeneous crystals have been used and different preform
manufacturing techniques have been applied, including mechanical
combination, preform growth, capillary drop and capillary suction.
Single-mode operation has been successfully demonstrated at a
wavelength of $10.6$\,$\mu$m for a fiber with a core/clad
composition of AgCl$_{75}$Br$_{25}$/AgCl$_{60}$Br$_{40}$, a core
diameter of $20$\,$\mu$m, a cladding diameter of $500$\,$\mu$m,
and a single-mode cutoff wavelength of $5.8$\,$\mu$m.
In a second activity, the team is currently developing single-mode
fibers by an improved and reproducible manufacturing process to
obtain fibers with improved performance. The test setup is
extended and allows interferometric measurements at wavelengths of
$5.3$\,$\mu$m, $10.6$\,$\mu$m and $16.5$\,$\mu$m to verify the
performance over almost the entire wavelength range. \noindent --
A team at Tel Aviv University (Israel) developed for NASA
silver-halide fibers for TPF-I. An improved crystal growth
technique allowed inhomogeneities of less than $2\%$ for preforms
of any composition.
An improved fiber has been developed with a double-step index profile. The core composition was AgCl$_{30}$Br$_{70}$, the composition of the first cladding AgCl$_{32}$Br$_{68}$ and that of the second cladding AgCl$_{5}$Br$_{95}$. The core diameter was $50$\,$\mu$m, the diameter of the first cladding $250$\,$\mu$m and that of the second cladding $900$\,$\mu$m (see Fig.~\ref{fig3} and \cite{shalem2005}). The outer cladding was blackened by exposure to UV radiation. \vskip 0.25cm

\subsubsection{Integrated Optics}\label{IO}

While Integrated Optics is a well mastered technology below 2\,$\mu$m, the situation is noticeably different at longer wavelengths. The manufacturing techniques need to be adapted to infrared materials with different structural, mechanical and thermal properties, like high fragility.
Possible technologies for manufacturing integrated optics are based on ion-exchange, waveguide chemical etching and photo-darkening.
\begin{figure}[t]
\centering\small
       \centering\includegraphics[width=4.32cm]{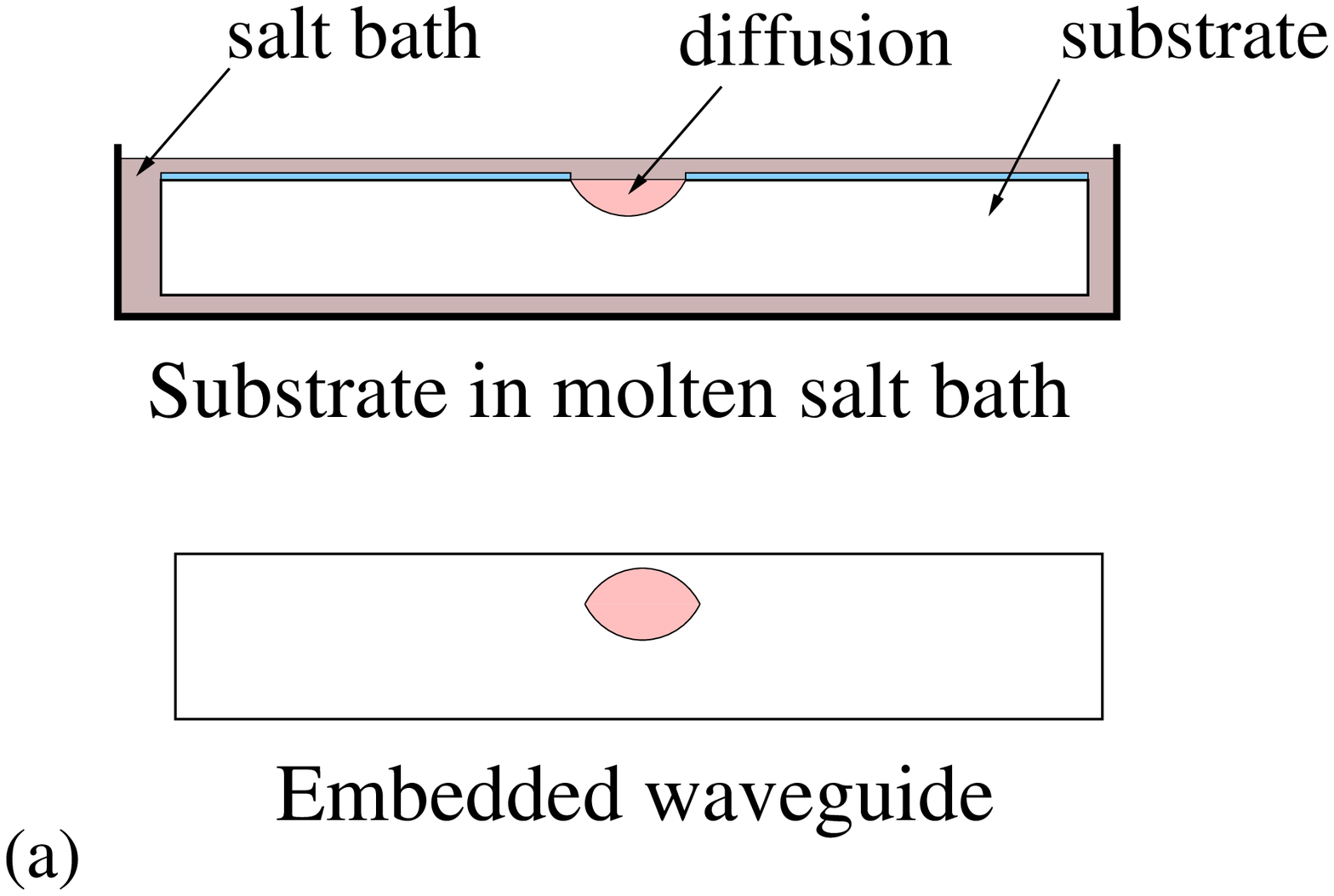}
       \hspace{0.0cm}
       \centering\includegraphics[width=4.32cm]{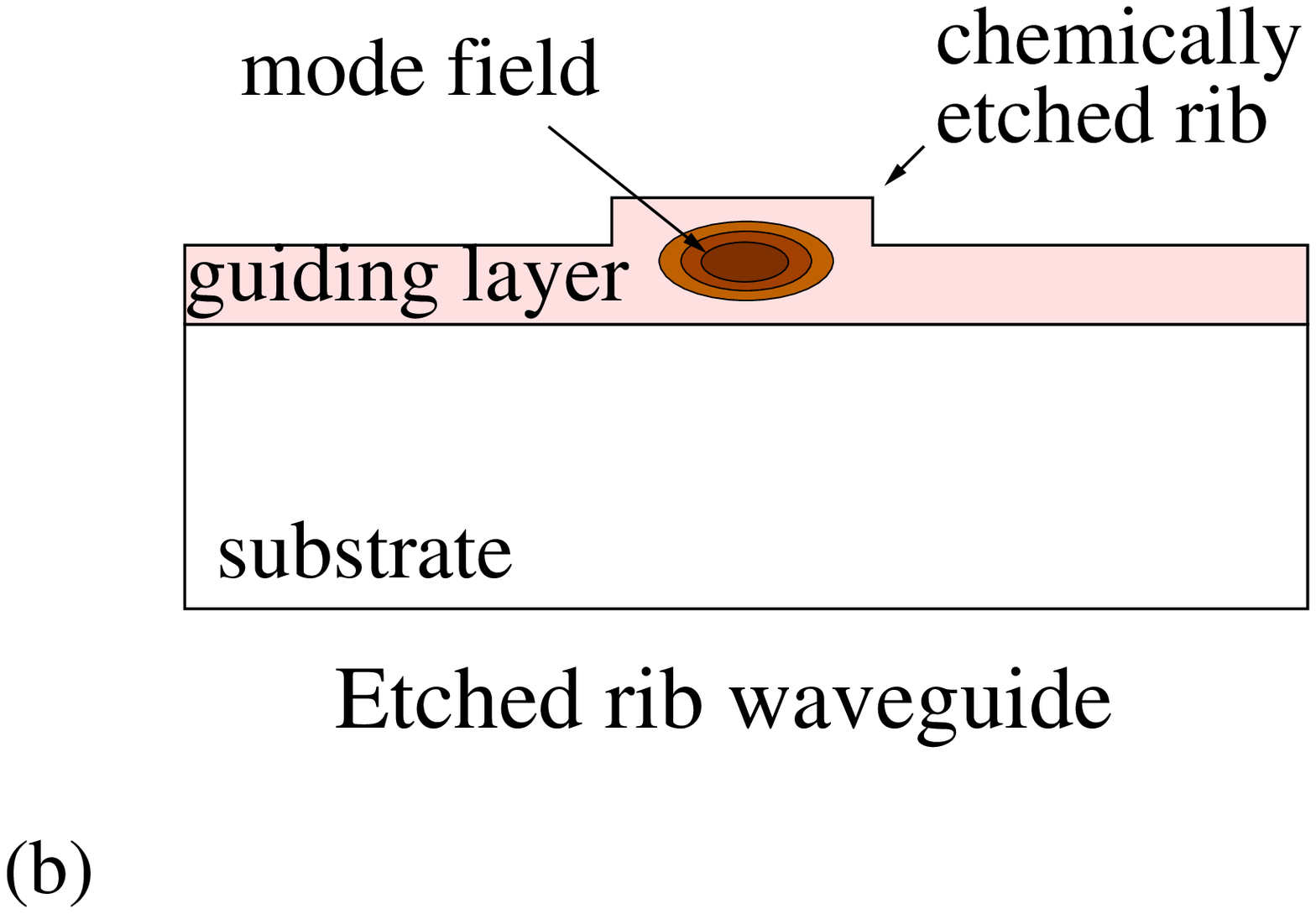}
    \hspace{0.0cm}
       \centering\includegraphics[width=4.32cm]{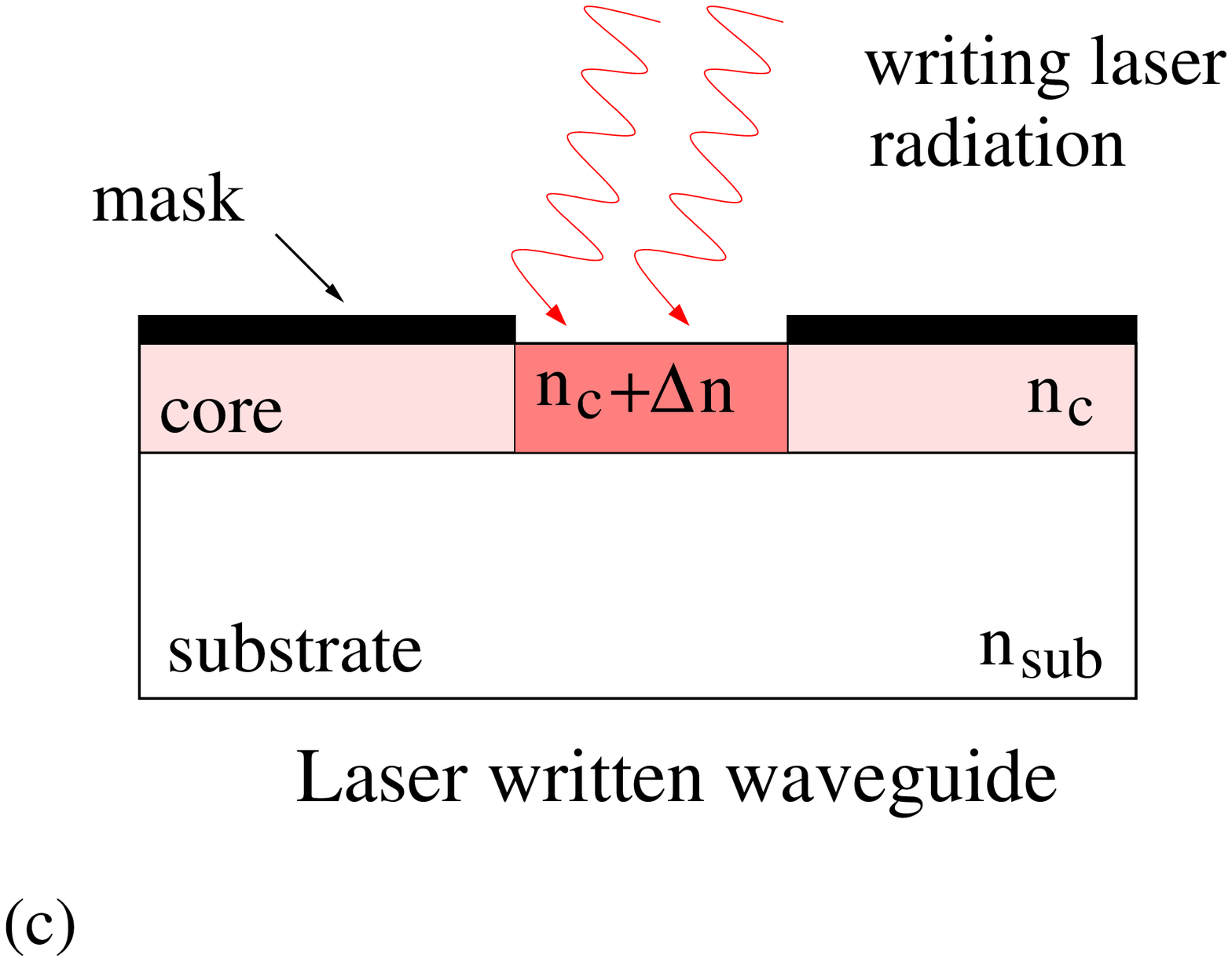}
\caption{Schematic view of the three main technologies for manufacturing integrated optics}\label{fig2}
\end{figure}
\paragraph{Ion-exchange diffusion} In the ion-exchange technique, a glass substrate is first layered with a few hundred nanometers of polysilicon coating with diffusion apertures obtained by ion etching (see Fig~\ref{fig2}(a)). The structure is then placed in a molten salt bath with precisely determined ion concentrations and treatment duration. The difference in concentration results in local ion exchange from the bath to the substrate, producing the high index core embedded in the substrate.
In the mid-infrared range, this technique has been successfully implemented on Germanate glasses for the 3--4\,$\mu$m spectral range, although the experimental characterization has so far been limited to a wavelength of 1.55\,$\mu$m \cite{Luo1998, Grelin2008}.
\paragraph{Chemical etching under controlled atmosphere} The chemical etching technique is among the most promising solutions to work with amorphous glasses like chalcogenide  (see Fig~\ref{fig2}(b)). Dry etching has already produced rib waveguides based As$_{\rm 2}$S$_{\rm 3}$, As$_{\rm 2}$Se$_{\rm 3}$ and Germanium components which were characterized at 1.5\,$\mu$m \cite{Ruan2004,Madden2007}. Later, analog rib waveguides were obtained using Tellurium compounds and characterized at 10.6\,$\mu$m (see Fig.~\ref{fig4} and \cite{Vigreux2007}). The first step in the manufacturing process is the deposition of a guiding layer by means of thermal evaporation or sputtering, followed by optical characterization \cite{Labadie2006}. These two processes influence the film parameters such as density, refractive index and surface roughness. A mask with the waveguide pattern is realized before undergoing chemical etching in Argon or CF$_{\rm 4}$/O$_{\rm 2}$ atmosphere. The process produces sharp, well defined waveguide contours, with step resolution better than 0.1\,$\mu$m. Some authors have demonstrated the feasibility of bent waveguides, which is an important prerequisite for more complex integrated optics functions \cite{Madden2007}. Chemical etching has been also attempted on crystalline glass compounds for slab waveguides made of a Zinc Selenide (ZnSe) film deposited on a Zinc Sulfide (ZnS) substrate \cite{LabadieThese}. However, the chemical and mechanical stability of the structure could not be preserved during the etching phase, and the layer was irremediably damaged.
\paragraph{Photo-darkening effect and laser writing}
A third way of producing mid-infrared integrated optics is to use the photo-darkening effect for amorphous glasses  (see Fig~\ref{fig2}(c)). Photo-darkening results from exposure of the substrate to high-power external radiation, in order to modify locally the refraction index of the glass. Although the correlation between photo-darkening effect and structural changes in the substrate is not well understood, it has been observed that photo-darkening occurs only in disordered amorphous materials, and not in crystalline materials. The process of waveguide manufacturing using the photo-darkening effect is named {\it laser writing}. This consists in focusing a visible source such as He-Ne laser in a particular location of the substrate surface leading to a change in the refractive index, which amplitude is a function of the exposure time to the radiation. This technique has been successfully applied to the As--S--Se family of chalcogenide glasses. Based on this technique, some authors could obtain a square cross-section single-mode waveguide with 5.4\,$\mu$m width and $\Delta$n$\sim$0.04 using laser writing at 632\,nm \cite{Ho2006}. This work has been done in the continuation of Efimov et al. (2001) effort to obtain analog results with 850\,nm laser writing. We wish also to underline that, depending on the exposition conditions, the photo-darkening effect can be reversible or permanent.

\begin{figure}[t]
\centering\small
\begin{tabular}{c c c}
 {\bf Single-mode silver } & \hspace{0.25cm} & {\bf Single-mode hollow}  \\
{\bf halide fiber} && {\bf metallic waveguide}  \\ [0.25cm]
\centering\includegraphics[width=3cm]{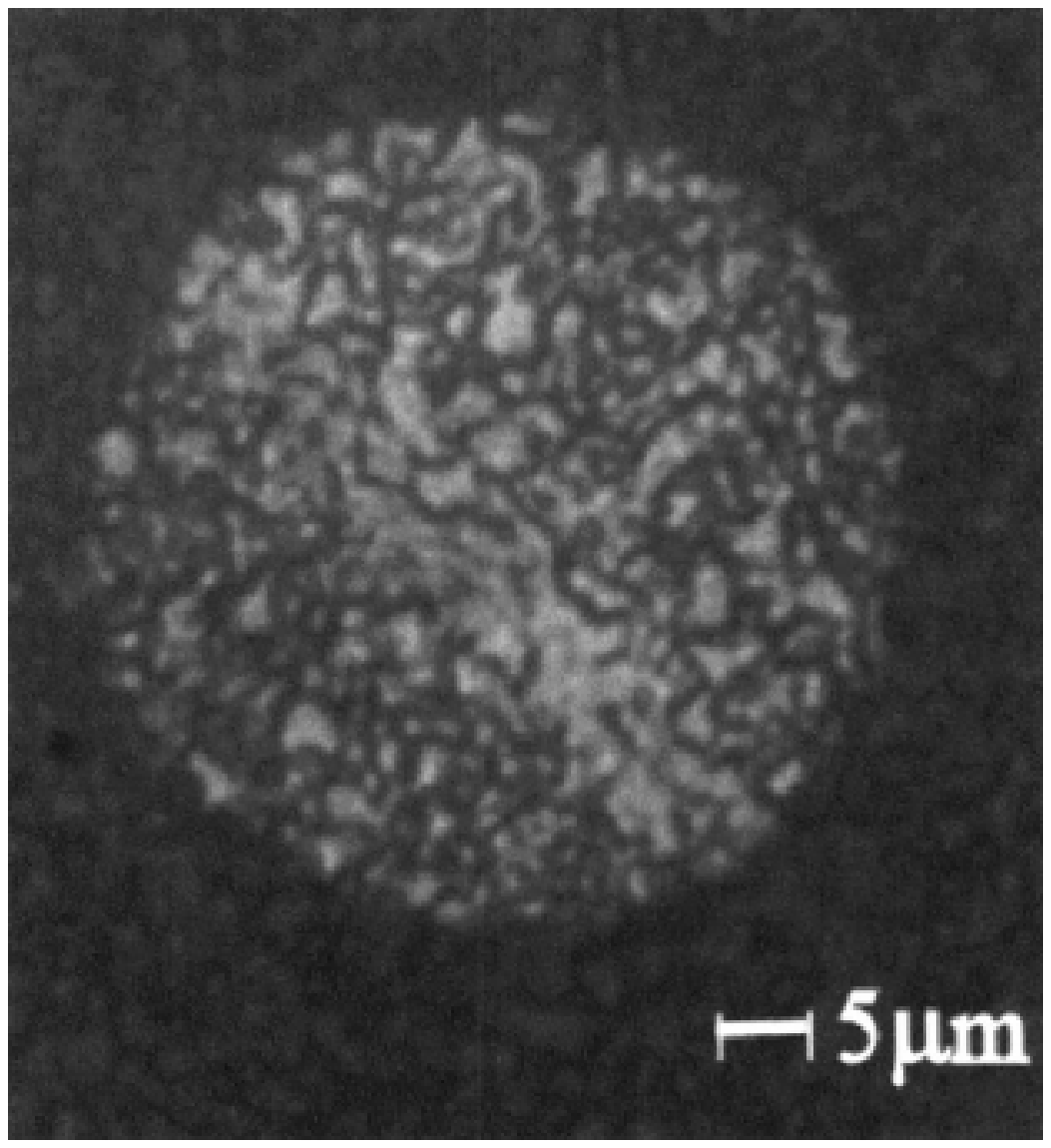} \hspace{4cm}&&
\centering\includegraphics[width=4cm]{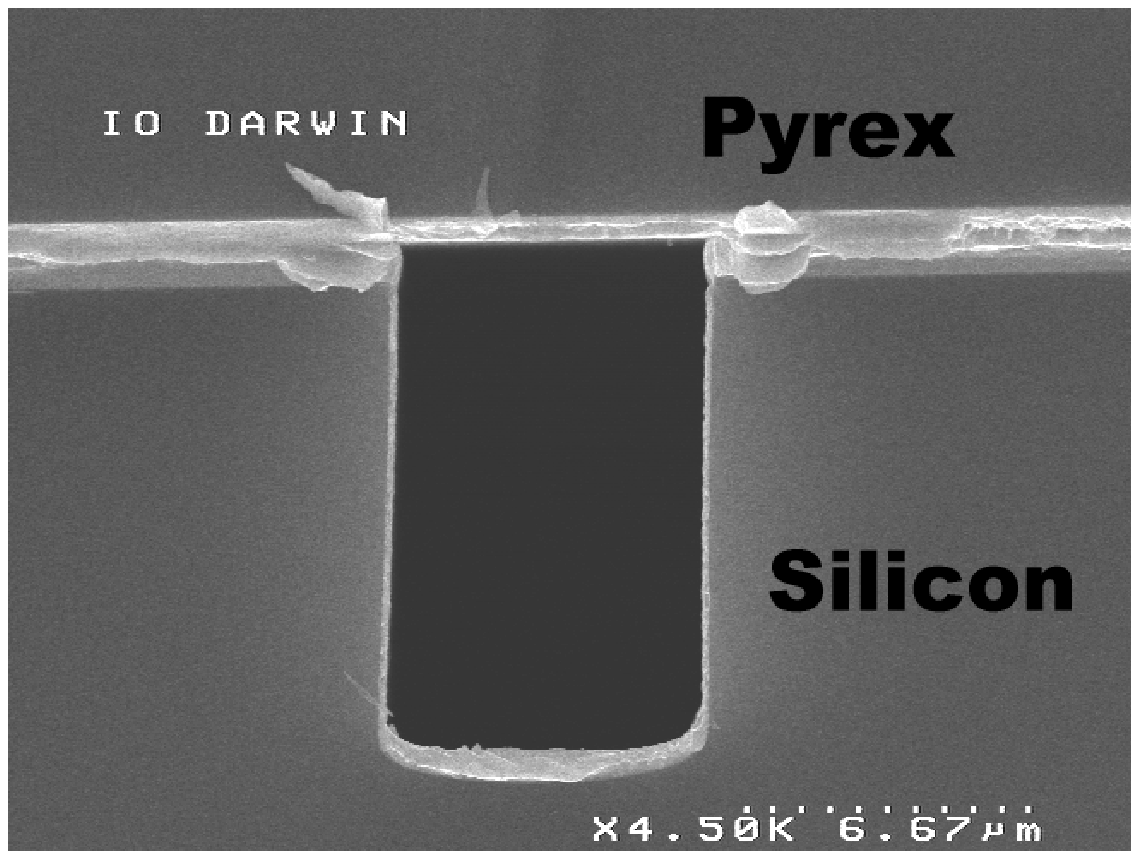}
\end{tabular}
\caption{Left:\, cross section of a single-mode silver halide fiber characterized by output profile imaging \cite{shalem2005}. Right:\, photograph of a conductive waveguide cross-section with dimensions 10$\times$5\,$\mu$m. This structure maintains a single polarization direction perpendicular to the longer side \cite{Labadie2006b}.}\label{fig3}
\end{figure}

\paragraph{Impact factors towards single-mode behavior} 
Beyond the manufacturing aspects of channel waveguides, the possibility to achieve single-mode behavior depends on the capabilities of manufacturing small cores (i.e. $<$10\,$\mu$m). This implies a higher refinement and mastering of the various technological parameters.
Alternatives to planar dielectric integrated optics exist to produce functions: a possibility is to emulate the near-infrared fiber couplers as this was done by Eyal et al. (1994) who mechanically joined uncladded 900-$\mu$m core silver-halide fibers to produce a Y-coupler for the mid-infrared \cite{Eyal1994}. Another option is to downgrade microwave E-plane and H-plane coupler to the 10-$\mu$m range as suggested by Wehmeier et al. (2004) \cite{Wehmeier2004}. However, obtaining a single-mode Y-coupler would require the handling of much smaller cores, while the solution of conductive waveguides needs additional developments to reduce the linear losses. Furthermore, the level of integration appears in both cases lower compared to planar integrated optics.

\subsubsection{Hollow waveguides}

While a certain amount of theoretical work has been produced to investigate the properties of {\it single-mode} hollow waveguides for the mid-infrared range, little has been done from the experimental approach.
Single-mode hollow waveguides have been limited by the actual micrometer capabilities of micro-machining techniques. Recently, different groups have addressed this manufacturability aspect. The group at Steward Observatory (Arizona) involved in radio and sub-mm instrumentation has tested the manufacturing of conductive waveguides by laser micromachining. This consists in etching a silicon wafer with high power (30 W) Argon laser in order to obtain a preform. These wafers are then gold coated, bonded and cut to obtain the waveguides. The accuracy and quality of the laser etching process is of crucial importance to realize the samples and minimize their losses. At the time of their published work \cite{Wehmeier2004}, the group could not achieve better than a 6\,$\mu$m resolution, which is compatible with operation in the 2-THz (150\,$\mu$m) regime. A second group based in Grenoble (France) has investigated a similar idea, but based on well mastered chemical etching process of the silicon wafer. The advantage is to obtain smaller structures with sizes down to 4\,$\mu$m and 50--100\,nm resolution, together with smoother waveguide walls. The deposited gold layer shows then a more uniform surface, which reduces scattering. The studied waveguide concept was inspired from well-known microwave rectangular waveguides, but scaled down to typical sizes of $\sim$10\,$\mu$m. Hollow metallic waveguides with rectangular geometry and with cross-section dimensions of 10$\times$5\,$\mu$m were produced with this technique and characterized at 10.6\,$\mu$m (see Fig.~\ref{fig3} and \cite{Labadie2006b}).
Later, Y-junctions based on hollow metallic waveguides were also manufactured, but with no sufficient transmission in the mid-infrared spectral range.\\

\begin{figure}[t]
\centering\small
       \centering\includegraphics[width=4.0cm]{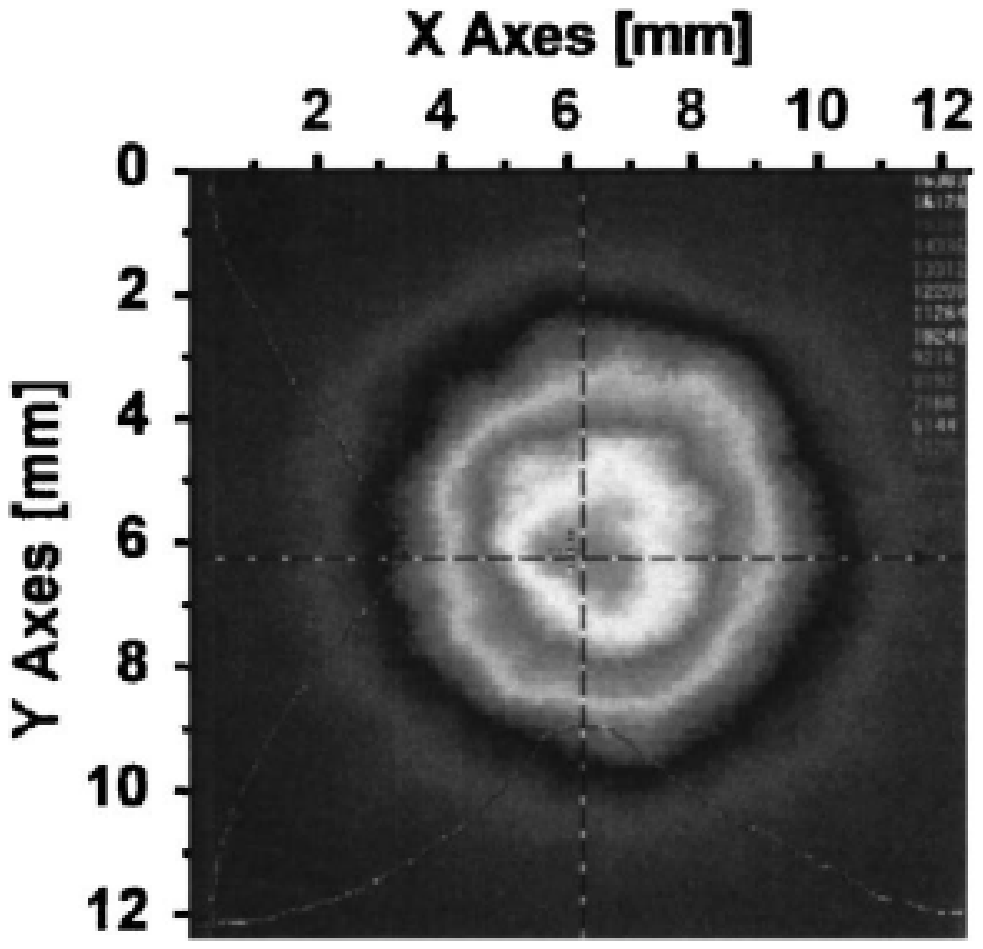}
       \hspace{0.5cm}
       \centering\includegraphics[width=6.0cm]{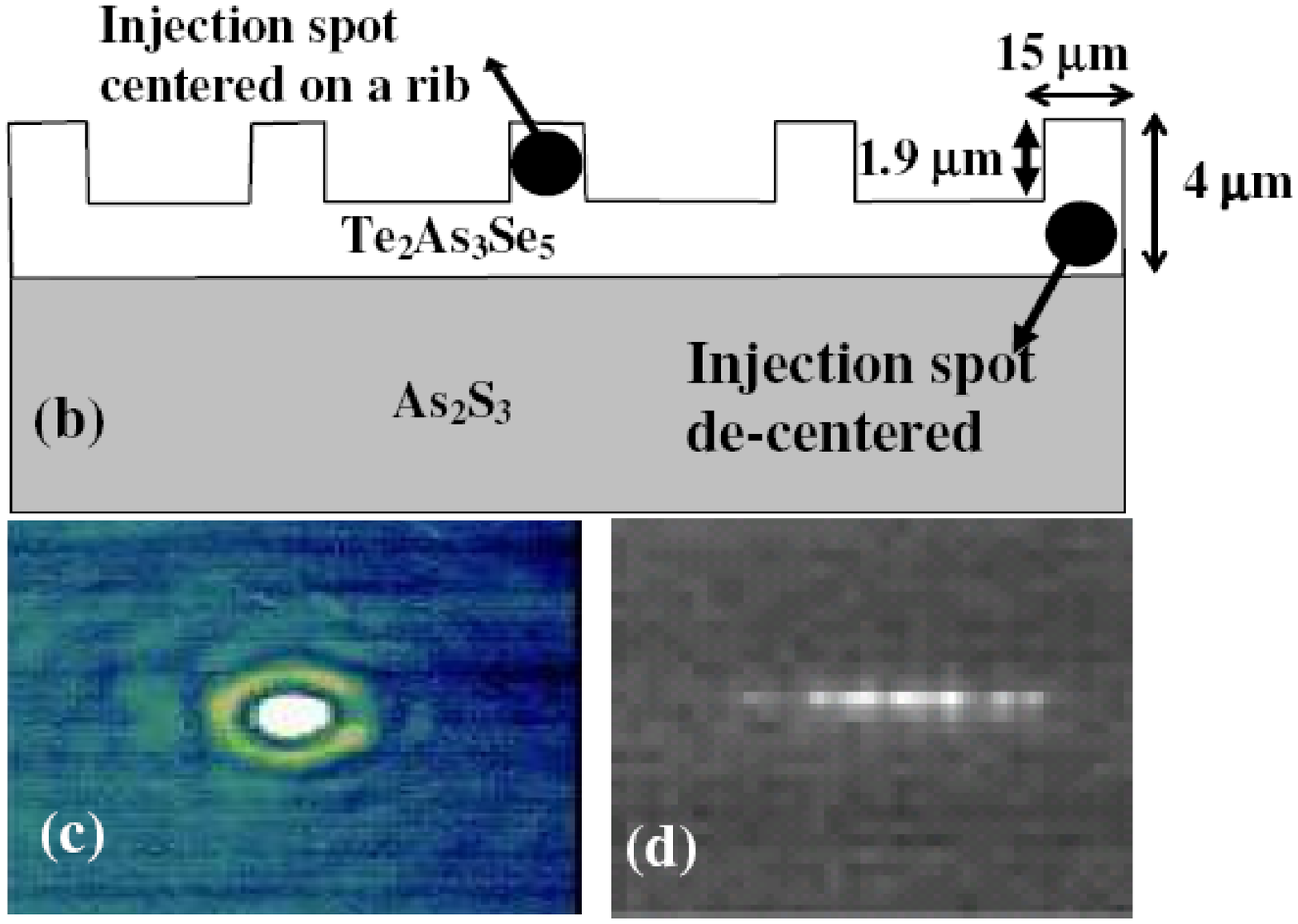}
\caption{Left:\, output profile of a 2.2-m long single-mode silver halide fiber obtained at 10.6\,$\mu$m \cite{shalem2005}. Right:\, sketch of a chalcogenide rib waveguide showing the position of the injection spot, together with output images at 10.6\,$\mu$m when the spot is centered and decentered from the rib \cite{Vigreux2007}.}\label{fig4}
\end{figure}

\subsubsection{Photonic crystal fibers}

The manufacturing techniques for photonic crystal fibers are similar to that for optical fibers and therefore are determined by the fiber materials. PCFs made of glassy material are drawn from the preform, PCFs made of crystalline material are extruded from the preform. The major difference with optical fibers is given by the preform fabrication technique. It shall be noted that PCFs for the mid-infrared are at a very experimental stage today.
For glassy PCFs the preform is usually obtained by drilling holes in the fiber material. For crystalline PCFs the preform cannot be obtained by drilling holes in the preform as these would be destroyed during the extrusion process. A reduced effective cladding index is obtained by adding a material of lower index to the cladding region \cite{astrium2007}. A one-dimensional photonic crystal is achieved by alternately coiling thin layers of silver-chloride and silver-bromide on a central silver-bromide rod. Two dimensional photonic crystals are achieved by filling the holes in a silver-bromide rod with small rods of silver-chloride, or by stacking silver-chloride and silver-bromide core-only fibers around a central rod of silver-bromide.\\
\indent Experimental PCFs made of silver-halide have been manufactured by several groups. The first index-guiding PCF was drawn
from a preform realized by stacking core/clad fibers around an uncladded AgBr fiber \cite{rave2003}. The fiber showed an effective
core diameter of $200$\,$\mu$m.
Other fibers have been realized from a silver-halide preform
made of AgCl$_{20}$Br$_{80}$ with holes drilled in a hexagonal pattern \cite{butvina2007}. The holes are filled with rods
made of AgCl$_{50}$Br$_{50}$. The extruded fiber showed a hole to distance spacing of $d/\Lambda=0.74$, an effective core
diameter of $79$\,$\mu$m and a numerical aperture of $0.16$.

\subsubsection{Considerations on design and technology}

In the context of astronomy, photonics devices face a certain number of requirements  linked to the original design of astronomical instruments, both from the design and technology point-of-view. For instance, classical optics (lens, mirrors...) used to feed an instrument are slow optics, with high numerical aperture that does not match one of fibers or integrated optics. This requires to use tapers, which in return can introduce additional losses if they are not adiabatic. Another point concern the importance of facet cleaning and anti-reflection coatings to limit Fresnel losses at the waveguide input. This is a technological aspect which is not yet very well accounted for, especially for mid-infrared waveguides. A trade-off between these different considerations is essential in the astronomical context, where transmission losses have to be kept as low as possible.

\subsection{Testing techniques and characterization results}\label{testing}

\subsubsection{Transmission range and excess losses}

Transmission range and losses are the two fundamental parameters
that define and fix the operability of the waveguides in the range
of interest. The combined effect of transmission range and losses
is very strongly dependent on the material, the applied
technological process and the waveguides geometry. This
measurement is the first step of any characterization phase.
\noindent -- In the case of dielectric waveguides, the
transmission range simply refers to the spectral domain where
photons are not absorbed by the material. Obviously, this
definition is linked to the transparency window of the bulk
material. The effective transmission range is provided through
spectroscopy measurements in the region of interest. For
dielectric bulks, the transparency window is obtained by FTIR
spectroscopy, which is a standard technique and will not be
further developed here. \noindent -- Excess losses regroup
propagation losses, coupling losses and Fresnel losses. Excess
losses are specific to each waveguide and directly depending on
its opto-geometrical parameters (length, numerical aperture, index
difference at air-waveguide interface). The different
contributions to excess losses are difficult to disentangle for a
single waveguide. Overall excess losses can be measured in
dedicated Fourier Transform spectroscopy experiments in which
light is launched into the waveguide placed in the optical path.
In some cases, it is possible to extract separately coupling and
propagation losses by differential measurements on similar
waveguides with different lengths and numerical aperture. This
method is also named ``cut-back method''. This method gives good
results, but requires then a dedicated set of waveguides planned
in the manufacturing phase \cite{Tiberini2007} \indent Measurement
campaigns on the different type of waveguides presented in
Sect.~\ref{manufacture} have shown a relatively large span in the
measured losses. Chalcogenide and silver-halides fibers show the
best transmission, even in single-mode regime, with losses at
10\,$\mu$m of 8 dB/m \cite{ksendzov2007}, 10 to 18 dB/m
\cite{cheng2005}, 20 dB/m \cite{borde2003}, 23.2 dB/m
\cite{flatscher2006}, 25 dB/m \cite{shalem2005}. Experimental
measurements at longer wavelengths for chalcogenide fibers have
led to increased values up to 40 dB/m between 12 and 13\,$\mu$m,
and 150 to 300 dB/m between 16 to 20\,$\mu$m \cite{faber2008}.
Dielectric integrated optics have presented so far higher losses
in the range of 1 to 10 dB/cm \cite{Vigreux2007, Ho2006}. However,
integrated optics is designed to operate over much shorter
propagation distances than fibers. At this stage of the
development of mid-IR waveguides, the reason of the high
dispersion in the propagation losses is mainly found in the
non-optimized manufacturing processes (impurities, defects
etc...).
More exotic waveguides like PCFs and hollow waveguide show
a quite different behavior. In two cases, mid-infrared
photonic-crystal fibers have shown very different losses of
$\sim$2 dB/m \cite{butvina2007} and 60 dB/m \cite{rave2003}, but
the single-mode behavior is not assessed. For hollow waveguides as
well, we found a strong dispersion in the performance. For most
multimode mid-infrared hollow waveguides, straight losses can be
as low as few dB/m, which is attractive for power transportation.
However, single-mode hollow waveguides are less advanced
technologically, and the first prototypes have shown high losses
of several dB/mm \cite{Labadie2006b}. Understanding such a
discrepancy requires further advances from the technological side.

\subsubsection{Single-mode regime and spatial filtering}

Supporting only a single spatial mode is a key feature of optical fibers for astronomical applications, and especially for mid-infrared interferometry \cite{clark1979}.
For mid-infrared fibers, one may think on different procedures
which rely either on measuring the fibers' far field intensity
distribution or on measuring the fibers waveguiding properties
(see Fig.~\ref{fig:measurement}). The procedures differ clearly in
complexity and accuracy:
\begin{figure}[t]
\centering\small
\begin{tabular}{ccc}
output beam profile & antisymmetric input field & nulling interferometer \\[0.2cm]
\centering\includegraphics[width=0.3\textwidth]{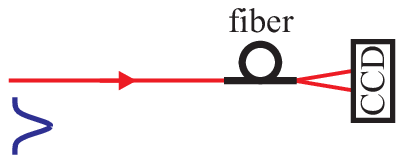} &
\centering\includegraphics[width=0.3\textwidth]{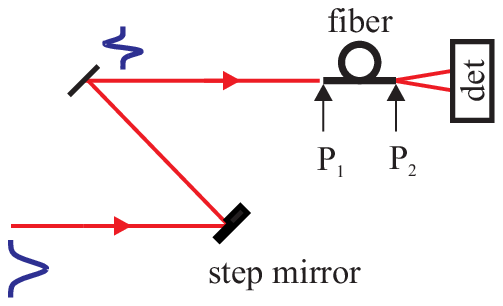} &
\centering\includegraphics[width=0.3\textwidth]{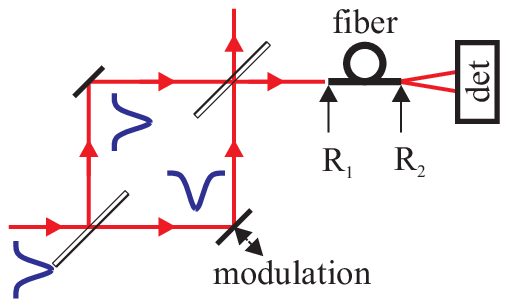}
\end{tabular}
\caption{Measurement procedures to verify single-mode behavior and spatial filtering of optical fibers. See text for details.
\label{fig:measurement}}
\end{figure}

\noindent -- The straightforward method is to measure the fiber
output beam profile at discrete wavelengths. If it corresponds to
the field distribution as expected from theory and if it does not
change in shape when changing the input beam launch conditions or
when bending the fiber, the fiber can be assumed to be
single-mode. The output beam profile can either be measured with
an infrared camera or by scanning with a single-pixel detector.
From the fiber divergence angle, the normalized frequency $V$ can
be directly deduced \cite{marcuse1981}. Due to the limited
accuracy, measuring the fiber output beam profile provides only a
qualitative assessment. \noindent --  Because the fundamental mode
of a single-mode fibers is symmetric with respect to the fiber
axis, the single-mode behavior and the spatial filtering
capabilities can be investigated by launching an antisymmetric
beam into the fiber \cite{ksendzov2007}. If the fiber is purely
single-mode, no signal can theoretically be measured at the fiber
output. Any leakage signal at the fiber output is due to
higher-order modes, back-reflection of cladding modes, or to a not
perfectly antisymmetric input field. The ratio of leakage power
$P_2$ to the incident power $P_1$ may serve as a qualitative
metric. A simple way for realizing an antisymmetric field is a
step-like mirror. The required antisymmetry of the input field is
the limiting factor for this method. \noindent -- Spatial
wavefront filtering with single-mode fibers is of particular
importance in nulling interferometry for equalizing the transverse
field distributions in order to achieve high rejection ratios.
Such interferometric test setup can therefore be used to verify
the single-mode behavior and the spatial filtering capability
\cite{wallner2004, Labadie2007}. For instance, the rejection ratio
(constructive to destructive interference power) of a Mach-Zehnder
interferometer is supposed to be much higher when employing a
single-mode fiber at its output ($R_2$) compared to the rejection
ratio achievable without the fiber ($R_1$). The ratio $R_2/R_1$,
also called filter action, serves as a quality metric for the
improvement in the rejection ratio. All measurements suffer from
residual light guided in the cladding. This is significant for
mid-infrared fibers, which have usually short length due to the
high attenuation. Absorbing coatings are therefore mandatory for
those astronomical applications that rely on the mode purity.
Furthermore, aperture masks may be applied at the fiber entrance
or exit. However, their achromaticity has to be taken into account
for broadband applications. \indent As an alternative to
monochromatic characterization methods, it is also possible to
detect modal cut-off wavelengths in the waveguide spectrum as less
power is transported with increasing wavelength. It is then
possible to identify in this way single-mode, bimode, trimode
regions etc... This method,  commonly applied in the near-infrared
\cite{Laurent2002}, is more delicate to implement in the
mid-infrared because of the lack of bright blackbody sources.
However, coupling the detection chain with a high-sensitivity MCT
single-pixel detector and a lock-in detection provides a
sufficient signal-to-noise gain to measure mid-infrared calibrated
spectra through the waveguide \cite{LabadieThese}. Low-loss
waveguides obviously help in increasing the experimental SNR.
Eventually, evidence of single-polarization in the case of hollow
rectangular waveguides can be exploited to demonstrate at a given
wavelength the single-mode nature of the waveguides. This method
has been successfully applied to assess the modal behavior at
10.6\,$\mu$m of the rectangular samples shown in Fig.~\ref{fig3}.
Among the mid-infrared waveguide prototypes presented in
Sect.~\ref{manufacture}, chalcogenide and silver-halide fibers had
single-mode behavior tested through output field imaging and
consequent qualitative fitting to the Gaussian profile of the
fiber fundamental mode (see Fig.~\ref{fig4} and
\cite{borde2003,cheng2005,shalem2005}). In the other cases,
nulling interferometry techniques were used to measure a high
order modes rejection ratio of about 1000$\pm$300
\cite{ksendzov2007} and a filter action $R_2/R_1$=400. The modal
behavior of integrated optics waveguides has been up to now
assessed by output field imaging.
In the last ten years, there has been a strong improvement in the measurement of the absolute rejection ratio obtained thanks to modal filtering through several experiments in the visible, near and mid-infrared. Several teams in Europe and in the U.S. have been working on nulling interferometry based on single-mode waveguides improving rejection ratios up to 10$^{4}$ to 10$^{6}$, depending on the wavelength range and bandwidth \cite{Haguenauer2006, Labadie2007, Buisset2007, Brachet2005, Weber2004, Flatscher2003}, and getting very close to the requirement of 10$^{5}$ broadband rejection ratio in the mid-infrared.
Note that other nulling experiments based on bulk optics setup provided very deep null, but they are not presented here since they do not have photonic devices implemented.


\section{Conclusions and perspectives}

We presented in this paper an overview of the research activities in the field of mid-infrared astrophotonics.  The effort from different groups has concentrated mainly on the development of single-mode waveguides like fibers and integrated optics. Many dielectric-based solutions explored for the mid-infrared are inspired from near-infrared concepts, like step-index and graded-index fibers. Other concepts like hollow waveguides, which are specifically designed to operate beyond 2--3\,$\mu$m and that do not have strong applications in the telecom bands, have been investigated with some success. Integrated optics at mid-infrared wavelengths is also emerging with the development of single-mode channel waveguides manufactured on planar waveguides. Photonics crystal fibers are part of a new generation of promising waveguides, but it should be acknowledged that PCFs for the mid-infrared are still at an experimental level. Single-mode behavior has been successfully demonstrated experimentally for different technologies, ultimately through nulling interferometry experiments where the measured quantity is the extinction ratio of the source light. Two hard points appear to be limiting factors for mid-infrared photonics: the first one concerns the high losses measured for these structures. However, this is mainly resulting from the non-optimized manufacturing techniques, which increasing maturity in the next years will help to improve the performance. The second point concerns the relative narrow bandwidth of the developed waveguides when compared to the astronomical needs. The experimental characterization has been often limited to a narrow wavelength range around 8.4\,$\mu$m or 10.6\,$\mu$m. Broadband characterization will help in the future to better answer to astronomical requirements. Although mid-infrared photonics has not reached yet the level of maturity obtained in near-infrared, the results of the last five to ten years are promising and open the way to new perspectives.\\
\indent Some concern the short-term improvement of the existing technologies. In particular, the demonstration of planar integrated channel waveguides is a key step towards first optical functions like beam combination. Furthermore, single-mode mid-infrared fibers have shown very good filtering properties in the context of nulling interferometry, which will be probably further improved in the coming years. In the long-term, while fibers appear as a rapidly maturing technology for mid-infrared stellar interferometry, integrated optics could be the ideal technology for instrumentation that requires to combine several optical functions. In particular, considering the future ELTs -- {\it Extremely Large Telescopes} -- integrated optics could have spectroscopic capabilities included within the same optical chip \cite{LeCoarer}, which is certainly of a major interest for facilities growing in size and cost. Eventually, beyond the frame of astrophotonics, mid-infrared photonics presents applications in the field of chemical and biological sensors, which makes them even more attractive for the next decades.

\section*{Acknowledgments}

The authors thank the anonymous referee for his useful comments to improve the scientific quality of the paper. The authors also thank Dr. T.\,M. Herbst for the constructive discussions and the accurate reading of the paper. LL gratefully acknowledges the funding support from the Max-Planck-Gesellschaft (MPG).

\end{document}